\documentclass[manuscript]{aastex}

\slugcomment{}

\shorttitle{Possible constraints on exoplanet magnetic fields}
\shortauthors{Scharf}

\begin{document}


\title{Possible constraints on exoplanet magnetic field strengths from planet-star interaction}


\author{Caleb A. Scharf}
\affil{Columbia Astrobiology Center, Columbia Astrophysics Laboratory, 550 West 120th St., New York, NY 10027}
\email{caleb@astro.columbia.edu}



\begin{abstract}
A small percentage of normal stars harbor giant planets that orbit within a few tenths of an astronomical unit. At such distances the potential exists for significant tidal and magnetic field interaction resulting in energy dissipation that may manifest as changes within the stellar corona. We examine the X-ray emission of stars hosting planets and find a positive correlation between X-ray luminosity and the projected mass of the most closely orbiting exoplanets. We investigate possible systematics and observational biases that could mimic or confuse this correlation but find no strong evidence for any, especially for planets more massive than $\sim 0.1$ M$_J$.
Luminosities and upper limits are consistent with the interpretation that there is a lower floor to stellar X-ray emission dependent on close-in planetary mass. Under the hypothesis that this is a consequence of planet-star magnetic field interaction, and energy dissipation, we estimate a possible field strength increase between planets of 1 and 10 M$_J$ of a factor $\sim 8$. Intriguingly, this is consistent with recent geodynamo scaling law predictions. The high-energy photon emission of planet-star systems may therefore provide unique access to the detailed magnetic, and hence geodynamic, properties of exoplanets. 

\end{abstract}





\section{Introduction}
Surveys for exoplanets have revealed a population of objects that orbit with very short periods (e.g. \citet{marcy2005}). The potential therefore exists for significant tidal and magnetic interaction between these planets and their host stars \citep{cuntz2000,ip2004,preusse2006,mcivor2006,cranmer2007}. 

Tidal evolution of planetary orbits, while still subject to significant uncertainty (due to the poorly constrained tidal quality factor $Q_p$, e.g. Matsumura et al. 2008), is well established (e.g. Jackson et al. 2008 and references therein). Tidal dissipation within planetary atmospheres is also a potentially important energy source in models of gas giant characteristics (e.g. Ibgui et al. 2009). In many cases the tidal evolution is dominated by the torques on the bulge raised on a planet by its parent star. However, for the most massive and closest orbiting planets, then significant torques are also possible due to the bulges raised in the stellar atmosphere. These modify the orbital evolution of the planet and act (for prograde planetary orbits) to maintain a higher observed spin rate for the host stars at a given age. Such an effect has been claimed in studies of exoplanet host star rotational velocities (Pont 2009), enhanced UV luminosities (Shkolnik et al. 2008) and at least one system - $\tau$-Bootis - shows evidence for stellar synchronicity with a planetary orbit \citep{walker2008}. 
It is also well established that main-sequence stellar X-ray luminosity is correlated with stellar rotation rates \citep{ayres1980,pallavicini1981,maggio1987} (see below), which are generally higher for the youngest objects.

Magnetic field interaction between stars and planets is likely complex, but is expected to exhibit distinctly different characteristics to tidally induced phenomena. In particular (and see below) we might expect stellar variations directly
related to planetary orbital phase and correlation with planetary mass and orbital parameters.
 There is growing evidence from the study of stellar photometry \citep{walker2008}, chromospheric emission \citep{shkolnik2008}, and X-ray emission \citep{kashyap2008, poppenhaeger2010} that interaction of this nature is indeed occurring between stars and planets - at least in certain cases. Furthermore, although the physical origin is not clear, \citet{hartman2010} finds a robust correlation between stellar chromospheric activity and the surface gravity of transiting hot Jupiters. 

X-ray, emission from the stellar corona is a particularly interesting probe of star-planet interaction. Coronal X-ray photons originate from thermal bremsstrahlung, and inner atomic transitions of the hot, $10^6-10^7$ K, plasma. The general heating mechanisms in stellar coronae are still largely unknown, but appear to be intimately related to magnetic field structures and complex transferral of energy from the stellar surface \citep{aschwanden2004,jess2009}. The placement of a planet, particularly a gas giant of Jupiter mass or greater, in proximity to the stellar coronal magnetic field may  effect the transport and release of energy in the plasma and therefore the high-energy photon output. Indeed, if stellar coronal magnetic field alignment favors strong coupling with planetary magnetospheres then significant, time-dependent, energy release may occur. Estimates suggest excess X-ray emission in the range of of $\sim 10^{27}$ erg s$^{-1}$ to as much as an order of magnitude variation in the quiescent stellar X-ray luminosity is possible \citep{ip2004,lanza2008,lanza2009b,cohen2009}). 

Stellar X-ray emission is of course determined by a variety of systematic and stochastic components, related to fundamental stellar activity (see above). The addition of an emission component that may vary on timescales commensurate with a planetary orbit poses a challenge for observations. At the distances for which both magnetic and tidal interactions are expected to become significant ($\sim 0.1-0.2$ AU) then orbital periods will range from $\sim 1$ to $\sim 20$ days (assuming stellar masses close to 1 M$_{\odot}$). Many X-ray pointed observations of nearby stars span only a few hours and may not adequately sample the possible planet-induced emission variation. Current techniques of planet detection \citep{otoole2009} are also significantly dependent on overall stellar activity, introducing further observational biases in sample selection.

 In order to further investigate the  possibility of coronal emission signatures correlated with close planetary companions we use a sample of exoplanet host stars within 60 pc with both X-ray detections and upper limits from the R{\"o}ntgensatellit (ROSAT) mission all-sky survey. We describe this data in \S2 and discuss its unique, partially time-averaged, characteristics which are well suited to this problem.  We examine the detailed relationship of X-ray luminosities and upper limits to planetary properties (\S 2.1, 3), and examine potential sources of bias (\S4). In \S 5 we consider tidal effects, and in \S6 we consider planet-star magnetic field interaction and the possible interpretation of our results. In \S 7 we discuss and summarize our results and conclude in \S 8.

\section{Exoplanet sample and X-ray data}

The ROSAT mission produced an all-sky survey of X-ray sources in the 0.1-2.4 keV energy band, released as a Bright Source Catalog (BCS, \citet{voges1999}) and a deeper, but less uniform Faint Source Catalog (FSC). 
In total $\sim 114,000$ sources were detected.

 Existing observations of planet-hosting stars suggest that X-ray emission triggered by a planet may be modulated over time, phase-shifted from the planetary orbital phase, and not necessarily in a precisely repeating pattern \citep{shkolnik2005,shkolnik2008,walker2008}.  Typical exposure times on our targets in the BCS and FCS ROSAT catalogs range from $\sim 400$ sec to $\sim 1000$ sec, however these represent accumulated exposure times from the $\sim 6$ month all-sky scan performed by ROSAT - with increasing exposure, and more scan passes, towards the ecliptic poles \citep{voges1999}.  The published source count rates therefore represent photons collected over a variety of timescales. At the ecliptic equator the $\sim 96$ minute ROSAT orbit (precessing by $\sim 1-1.5^{\circ}$ per day) allowed the 2-degree FOV of the detectors to scan across a source $\sim 20-30$ times over a 2 day period, with a maximum of 30 sec exposure in any given pass. At higher ecliptic latitudes (and therefore higher net exposures) this number of passes clearly increased towards a maximum at the ecliptic poles - which were in effect sampled every 96 minutes for 6 months. The use of the ROSAT all-sky data therefore directly complements pointed observations that may only sample a small fraction of a planetary orbital phase (e.g. \citet{kashyap2008,poppenhaeger2010}), and provides a time-averaged estimate of X-ray luminosity over a significant fraction of a planetary orbit, or even over multiple orbits at high ecliptic latitude - albeit with significant error bars.
 
We cross-correlate the BCS and FCS source catalogs with a sample of 271 exoplanet host stars drawn from the online repository of all exoplanet detections at the time of this work \citep{schneider2010}. The exoplanet host stars are chosen to include only systems with determined orbital parameters (semi-major axis and eccentricity) and known system distances. As described below, we further trim the data to exclude giant or sub-giant stars. 

We use NASA's HEASARC Browse archive interface to perform an automated batch search for X-ray counterparts and set the search radius to 0.6 arcmin, adopting the same criteria employed by other efforts to assign stellar X-ray luminosities to stellar targets using the ROSAT All Sky Survey (RASS) data (e.g. \citet{kashyap2008}). Beyond search radii of 0.66 arcmin, source identification is subject to significant confusion \citep{voges1999}.  Although all-sky, the ROSAT source catalogs have varying detection sensitivity according to net exposure time during the survey (see above). The bright (BCS) sample has 92\% sky coverage to a background subtracted count rate of 0.1 ct s$^{-1}$ (0.1-2.4 keV) or $6.5\times10^{-13}  $erg s$^{-1}$cm$^{-2}$ in flux for a 0.5 keV thermal spectrum broadly consistent with stellar thermal coronal emission. The Faint Source Catalogue (FSC) sample detection limit is effectively based on a net photon count of 6 \citep{voges2000}. The faintest counterparts to our stellar catalog have a background-subtracted count rate of $0.013\pm0.006$ ct$^{-1}$ in the FSC. Count rate errors are taken from the survey and include errors in the estimated background rate at the source location.

 Distances to all stellar objects are obtained by cross-correlating with the Hipparcos astrometric catalog, again through NASA's HEASARC Browse facility. A search radius of 6 arcsec is used, which was found to result in unique identifications in all cases here. Errors in parallaxes were not propagated through to the final errors in X-ray luminosity. Another potential source of uncertainty arises from flux contributions by physically associated stellar companions. At least $\sim 20$\% of known exoplanet host stars have stellar companions \citep{raghavan2006} whose emission could confuse and/or contaminate the emission from the planet hosting star. The RASS data, and most pointed X-ray data, lack the spatial resolution to deconvolve such contamination. False positive detections due to background (typically AGN) X-ray sources is estimated to be low - for the search radius of 0.6 arcmin around 271 planet-host stars we would  expect (based on the RASS BSC and FSC source catalog sky densities) a total of only $\sim 0.2$  false detections.

We obtain 35 unique X-ray detections of planet harboring systems. Excluding giant or sub-giant stars results in a final sample of 29 systems of main-sequence stars, all within 60 pc distance (shown in Table 1). Twenty-two of these stars are K or G spectroscopic types, two are M dwarfs and 5 are F stars, masses range from 0.35 to 1.28 solar masses, with a mean of 0.93 solar masses. X-ray luminosities are computed from photon counts assuming a conversion of  $6.5 \times 10^{-12}$ ergs cm$^{-2}$ ct$^{-1}$ appropriate for the ROSAT 0.1-2.4 keV band and a coronal thermal spectrum (e.g. \citet{kashyap2008}).

We also estimate upper limits to X-ray luminosities for the planet harboring systems with no detected X-ray counterpart.  At the time of this analysis the total number of systems used is 90 with planet periastron distance within $0.15$ AU, and 145 with planets whose periastron distance is greater than or equal to $0.15$ AU (see \S3 below). Background count rates per arcmin$^2$ estimated from the RASS source detection algorithm (0.1-2.4 keV) were extracted from the RASS database for all source detections within 60 arcmin radius of the non-detected system coordinates. These data contain no less than 3 detections, and a mean of 17.6 detections for all systems. The mean background rate per arcmin$^2$ ($\bar{CR_b}$) for these surrounding sources was computed and used as the basis for estimating the non-detection upper limit (using the exposure time at the stellar target coordinates, $t$).  BSC and FSC RASS count rates for point sources are given within a smallest size aperture of 5 arcmin radius \citep{voges1999} and we use this as our default extent for estimating a 5-$\sigma$ count fluctuation as our upper limit (i.e. Poisson statistics are assumed and the upper limit count rate is: $(5 \sqrt{\bar{CR_b} \times t \times \pi 5^2})/t$ ). Count rates are converted to luminosities as described above.

Finally, we impose a volume limit to the entire sample at 60 pc. All X-ray detected systems lie within 55 pc. A significant fraction of the systems with upper limits are the result of transit survey detections (which is strongly biased to short period objects due to the geometrical detection probability) at distances of over 100 pc, with subsequent radial velocity followup. While imposing a volume limit is not ideal, it does effectively remove this highly biased population. The final conclusions reached here are not significantly altered by the effect of this distance cut. Our resulting set of X-ray upper limits is then reduced to 42 systems with periastron within $0.15$ AU and 99 systems with periastron greater than 0.15 AU.

\subsection{Inner and outer planet subsamples}

To divide the planet-hosting stellar sample between those with a high potential for magnetic field or tidal interaction and those without, we evaluate the periastron distance ($d_{peri}$) of the innermost planet (Table 1). An inner subsample is defined for systems where $d_{peri}<0.15$ AU. This choice of distance is commensurate with that used in earlier studies \citep{kashyap2008} and represents an orbital period of $\sim 20$ days around a one solar-mass star, within which tidal interactions between star and planet become dynamically important \citep{goldreich1966} and orbital velocities are similar to coronal field rotation velocities (\citet{lanza2009b}, and below). Our conclusions are unaltered for variations in this distance of $\pm 0.05$ AU. 

Our statistical examinations of these data are made using the survival analysis methods of the ASURV version 1.2 software package \citep{isobe1990,lavalley1992}, that allows for `censoring'  (i.e. upper limits) and univariate and bivariate tests by implementing the methods presented in \citet{feigelson1985} and \citet{isobe1986}. Thus, both detections and upper limits are incorporated into all statistical tests in a self-consistent and robust fashion.

In Table 2 we present the results of a Kaplan-Meier (K-M) estimation of the underlying distribution of the inner and outer sample X-ray luminosities. The two samples agree well in all but the 75th percentile of the low $L_x$ end of the distributions, where there is a difference of $\Delta \log_{10} \sim 6.5$. This is directly attributable to the lowest $L_x$ {\em detections} occurring in the inner sample only. As we discuss further below, this may be due to the observational biases of the RV planet detection method, wherein the very least massive planets are only detectable on close orbits around low-activity stars. In all other respects the inner and outer samples show statistically similar $L_x$ distributions. The means differ by approximately one standard deviation. We caution though that K-M means and medians are not reliable for very heavily censored data, and that our present samples include significant numbers of censored points (upper limits). Nonetheless, these results are in general agreement with the work of \citet{poppenhaeger2010} who find no excess emission from planet-hosting stars in general. Our results do not support the findings of \cite{kashyap2008}, who claim as much as a factor 4 emission enhancement.

We also apply tests to determine whether the two censored populations are drawn from the same underlying distribution. A Gehan generalized Wilcoxon test (that assumes the same censoring pattern applies to both samples being compared) yields probabilities of 91.99\% (permutation variance) and 92.11\% (hypergeometric variance, more robust to censoring pattern differences) that the $L_x$ detections and upper limits of the inner and outer planetary samples are indeed drawn from the same physical distribution.

In summary; there does not appear to be any significant difference in the $L_x$ distributions of the inner and outer subsamples, especially if we allow for observational biases that particularly affect the lowest mass planet detections.

\section{Planet mass vs $L_x$ correlations}

We examine the detailed relationship between measured planetary properties and system X-ray emission in our sample. In Figure 1, detected X-ray luminosity is plotted versus the radial velocity (RV) derived planetary mass ($M_p\sin i$, projected by the typically unknown system inclination) for the outer ($d_{peri}>0.15$ AU) and inner ($d_{peri}<0.15$ AU) planetary subsamples. The presence of the lowest mass planets in only the inner sample is  a result of biases in planet detection techniques. For a given instrumental RV sensitivity then the lowest mass planets are only detectable on short orbits. Furthermore, the least active stars are favored for the detection of small RV variations and lower mass planets. In \S4.1 we examine the residual stellar `jitter' in our sample systems and this bias in more detail. However, there is {\em also} a clear difference between the distributions seen for planets with $M_p\sin i>0.1$ (M$_J$) between the inner and outer subsamples  that is not readily explained by observational biases. $L_x$ shows a correlation with $M_p\sin i$ for the inner subsample that is not evident in the outer subsample. 

In Figure 2, we plot detected X-ray luminosities {\em and} upper limits versus projected planet mass for the inner and outer subsamples. For the inner planet subsample the distribution of upper limits appears to be consistent with a population of sources that either follows the power-law fit seen in detected systems or has a lower bound or `floor' that correlates with planet mass ($M_p\sin i$). A notable outlier in the upper limits of this subsample is labeled as `a' in the inner planet panel of Figure 2. This point corresponds to the GL 86 system at an 11 pc distance, with a single $\sim 4$ M$_J$ planet at 0.11 AU semi-major axis and eccentricity $e=0.046$. GL 86 A is a K0V star in a likely binary configuration with a 0.5 M$_{\odot}$ {\em white dwarf} companion (GL 86 B) at $\sim 20$AU separation  \citep{lagrange2006}. Furthermore, there is a potential X-ray detection at 0.63 arcmin separation, just beyond our search radius. If that source is at the distance of GL 86 it has $L_x \sim 6 (\pm 1) \times 10^{27}$ erg s$^{-1}$, commensurate with our estimated upper limit at that location. The history of this system is somewhat uncertain, and the presumed earlier presence of a $\sim 2$ M$_{\odot}$ GL 86 B stellar progenitor \citep{lagrange2006} at smaller separation raises considerable uncertainty about the environmental history of the planet GL 86 b and its parent star.

Applying the tools of survival analysis we find that for the outer sample a generalized Kendall statistic yields a 26.4\% probability that {\em no} correlation is present between $L_x$ and $M_p \sin i$. A generalized Spearmans $\rho$ test yields a 11.3 \% probability of no correlation. By contrast, for the inner planet sample the generalized Kendall strongly indicates the presence of a correlation, with only 0.01\% probability that there is none, and 0.09\% from the Spearmans test.

We fit for a power law correlation in both samples using a Buckley-James (BJ) regression with Kaplan-Meier residuals (a more restrictive EM algorithm fit yields essentially identical results in all cases), and the results are summarized in Table 3 and Figure 2. These regressions make full use of both the detected and censored (upper limit) data. The outer planet sample fit is consistent with no correlation, the inner sample shows a very significant correlation. In Table 3 we also show the regression analysis results when the X-ray upper limit for system GL 86 is removed from the inner subsample - the change to the correlation is negligible. Our final best fit (including the GL 86 upper limit) to the correlation seen in the inner planet subsample is: $L_x=1.64  \times 10^{28}\; (M_p\sin i)^{0.59\pm0.04}$ erg s$^{-1}$ ($1\sigma$ errors), using the BJ regression. We also note that these results are largely unchanged if we relax the 60 pc distance limit for the sample to include all 271 exoplanet systems (\S 2), or if we exclude systems with $M_p\sin i <0.1$M$_J$.

\section{Observational biases}

Although a clear, statistically significant, correlation is measured between X-ray luminosity and projected planetary mass for the inner planet subsample - but not for the outer sample - the nature of planet detection techniques is such that care must be taken to eliminate the possibility of selection biases creating an apparent correlation between observed properties. In particular, RV detection sensitivity is affected by stellar activity. It  might also be possible that planet populations at different orbital distances are influenced by the physical environment - giving rise to apparent correlations between planetary and stellar characteristics. This is a recognized problem (e.g. Hartman 2010) and is difficult to resolve. We have made a preliminary investigation of possible biases, described below.

\subsection{Jitter and inclination}

We first examine the relationship of the known uncertainties in RV measurements used for planet detection and the measured X-ray luminosity of the stars in our planetary system sample.  Residual errors in velocities following the successful fitting of planetary companion orbits are generally reported as r.m.s. values. These implicitly include both the systematic instrumental errors (typically a few m s$^{-1}$) and the intrinsic uncertainties due to stellar ÒjitterÓ, a consequence of the finite distribution of velocities in the stellar photosphere due, for example, to convection. We have taken r.m.s. values from the primary publications for the most up-to-date RV measurements, quoted in the online Extrasolar Planets Encyclopaedia \citep{schneider2010}.
In Figure 3 the detected X-ray luminosity is plotted versus the r.m.s. stellar jitter for our complete exoplanet sample, and for just the high-mass ($>0.1$ M$_J$) inner subsample. As might be expected, the lowest mass planet detections correspond to the lowest r.m.s. errors. There is however no apparent correlation between r.m.s. errors in radial velocity and stellar X-ray emission detection  for the more massive ($>0.1$ M$_J$) subsample. This suggests that any systematic bias in planet detection introduced by stellar activity (i.e. the preferential detection of only higher-mass planets around more active and hence X-ray luminous stars) should be negligible in the bulk of our sample.  A potential correlation, or grouping, is seen for the 4 systems with $M_p\sin i<0.1$ M$_J$, which is not unexpected given that the detection of these planets is presently only feasible for the very least active stars. However, any general observational biases against the detection of lower mass planets around more active stars might be expected to affect larger mass planets in the outer sample as well,  given the reduced RV signal amplitude with increasing planet-star separation (scaling as $d^{-1/2}$). No such bias is apparent. 

The jitter estimates are themselves subject to significant variation. This originates from the specific instruments used to obtain RV spectra (of which there are many, and many `upgrades' spanning the past 15 years of RV planet detection), observing conditions and sampling strategies, and the analysis tools used to obtain the best fit planetary parameters to the RV curves.

Geometry biases the RV detection of planets to low inclinations with respect to the observer. If coronal X-ray emission due to magnetic field interactions occurs preferentially in localized regions directly between the planet and star \citep{cuntz2000,mcivor2006,cohen2009} this might introduce a systematic trend for higher X-ray emission visibility at low inclinations, which could be correlated with systematically larger estimated planet mass (i.e. larger $\sin i$). The presence of this bias would of course itself be evidence for planet-star interaction.
This could be tested using a population of transiting systems, where the inclination is known and typically close to $\sim 90^{\circ}$, however in the present sample of 29 X-ray detected systems only one (HD 189733) is transiting.

\subsection{Stellar rotation}

It is well established that main-sequence stellar X-ray luminosity is correlated with stellar rotation rates \citep{ayres1980,pallavicini1981,maggio1987} and   $L_x \approx 10^{27} (v_*\sin i)^2 $ erg s$^{-1}$ (where $v_*\sin i$ is in km s$^{1}$) up to a point of saturation at $L_x/L_{bol}\approx 10^{-3}$. Physically, for slow and intermediate rotation, this relationship has been cast in terms of the Rossby number: $R_0=P/\tau_c$, where $P$ is rotation period and $\tau_c$ is the convective turnover time (e.g. see \citet{gudel2004} and references therein). Saturation then corresponds to $R_0 < 0.1$, but which rotation rate this corresponds to decreases with decreasing stellar mass \citep{pizzolato2003}. For a  1.05 M$_{odot}$ star it is $P\approx 1.5$ days, for a 0.7 M$_{\odot}$ star it is $P\approx 3.5$ days \citep{gudel2004}. None of the stars in our planet sample appear to reach saturation with $L_x/L_{bol}\approx 10^{-3}$. Rotation rate is also known to correlate with stellar age - with younger systems exhibiting shorter periods. We note that the three systems with highest $L_x$ in the inner planet subsample (see also Figure 6) span a range of ages: $\tau$-Boo: 2.5 Gyrs, HD 162020: 0.76 Gyr, and HD 41004B: $1.6\pm 0.8$ Gyr (values taken from the literature), and are not exceptionally young.

 In Figure 4 we plot the available $v_*\sin i$ estimates (obtained from the primary data references given in the online exoplanet catalog \citep{schneider2010}, or alternatively if not available there then from the primary reference given in SIMBAD) versus the $M_p \sin i$ for both our inner and outer planet X-ray detected subsamples. We note that the stellar inclination need not be the same as that of the planet orbit. As seen with the stellar jitter (Figure 3) the datapoints corresponding to the lowest mass innermost planets ($<0.1$ M$_J$) appear to group with a lower mean $v_*\sin i$ than the more massive planet detections. This is consistent with the lower mass planets being more readily detectable in RV data with less line broadening or activity. 
It therefore seems that the lower X-ray luminosities of the four stars hosting the least massive innermost planets is consistent with the lower rotation rate of these stars, however it is not apparent that this trend is present in the larger mass subset.

At higher masses ($>0.1$ M$_J$) there is no strong correlation, however we note that  there may be a division between faster rotating and slower rotating stars. At masses above $\sim 3$ M$_{J}$ there may be two groupings of $v_*\sin i$ - one high at $\sim 10$ km s$^{-1}$ and one low at $\sim 2-3 $ km s$^{-1}$.
 \citet{pont2009} has recently discussed issues of tidal evolution in close-orbit systems, including the possibility of evolution towards synchronization of the stellar photosphere for $M_p >\sim 3$M$_{J}$. It is possible that this present data reflects some of these effects, in the form of `excess' stellar spin due to planetary companions.

We further examine the relationship of system X-ray luminosity estimated from the empirical relation $L_x\approx 10^{27}(v_*\sin i)^2$ ergs s$^{-1}$ (see above) to that observed (Figure 5). The original $L_x\approx 10^{27} (v_*\sin i)^2$ relation was derived in the {\sl Einstein} observatory band of 0.2-4 keV. We have therefore converted the predicted luminosities to the 0.1-2.4 keV ROSAT band assuming the same thermal coronal model as used to convert counts to flux (\S 2) and find $L_x\;(0.1-2.4) \approx 1.31 \times 10^{27} (v_*\sin i)^2$ erg s$^{-1}$.

While there is a general correspondence between the estimated and measured luminosities there is very significant scatter, and for the outer exoplanet subsample there appears to be a systematic underestimate of $L_x$ from stellar rotations. 

\subsection{Planetary erosion}

There is evidence in the literature that suggests the population of planets in short orbits may have been modified by atmospheric erosion (e.g. \citet{lammer2003}). \citet{sanz-forcada2010} examine this using the X-ray {\em fluxes} at the location of planets in a sample of 59 systems with {\sl XMM}, {\sl Chandra}, and ROSAT X-ray data and claim an excess of lower mass planets in environments of high flux. One explanation could be that atmospheric erosion by high-energy photons has been so efficient that only eroded planets are seen in high flux locations. Other results, examining the relationship of chromospheric activity (via the strengths of Ca II H and K lines) to the surface gravitational acceleration computed for transiting planets do not support this conclusion \citep{hartman2010}.

However, while erosion may occur, it does not indicate a population bias that would produce the effect seen in our analyses - where the most massive, most closely orbiting  planets correspond to the highest coronal emission luminosities (see also Figure 6 below). It is possible that a further correlation with system age (whereby old systems preferentially harbor the most eroded planets) could reconcile this bias with our results - however (\S 4.2) there is no obvious age-related correlation in our sample. We also note that the majority of X-ray luminosities used by \citet{sanz-forcada2010} are derived from pointed X-ray data from {\sl Chandra} and {\sl XMM} with exposure times typically of the order of 10 ksec (and do not appear to exclude sub-giant stars, e.g. $\gamma$ Cephei ). As described in \S2, such data typically sample only a fraction of orbital phase compared to the ROSAT all-sky survey, and so while entirely complementary there may be biases introduced that preclude direct comparison.

\section{Tidal effects}
For completeness we also examine the relationship between X-ray emission and a scaling parameter proportional to the height of the tidal bulge raised in the stellar atmosphere by a planet at its periastron distance: $M_p/d_{peri}^3$. Projected planet mass is used here. Figure 6 illustrates this distribution. The three systems (labeled) with the largest tidal effect correspond to the three most massive planetary objects in our inner sample, with high $L_x$, suggesting tidal interaction could also play a role. This is broadly consisent with the findings of \citet{pont2009}, where evidence was presented for `excess' stellar rotation due to star-planet tidal torquing in the most extreme cases. These would appear to be excellent targets for future efforts to investigate star-planet interactions.

\section{Interpretation and physical models}

The X-ray luminosities of main sequence stars containing closely orbiting ($d_{peri}<0.15$ AU) planets appear to be correlated with the lower limit to planetary mass. The upper limits on X-ray luminosities for planet-harboring systems without X-ray detections are also consistent with this correlation. There is no evidence for this correlation in systems with more distant ($d_{peri}>0.15$ AU) planets, suggesting that observational selection effects (that might impact low amplitude RV detections of low mass planets on short orbits as well as larger mass planets on long orbits) cannot be entirely responsible.

Assuming that this is a consequence of the physical interaction of these planets with their parent stars, then there are two potential forms of the observed correlation. The first is that the X-ray luminosity is entirely governed by the planet-star interaction (which can be excluded on purely energetic and physical grounds, e.g. \citet{lanza2009b}), the second is that the X-ray luminosity is a combination of the underlying, or `normal', coronal emission, and an additional component due to the planet-star interaction that effectively produces a lower-limit floor to the distribution of luminosities - i.e. a minimum X-ray luminosity that is a function of planet mass. 
The lack of any statistical evidence for a difference in the mean $L_x$ between our inner and outer subsamples is not inconsistent with planet-star interaction causing a minimum $L_x$ as a function of planet mass. The large range in $L_x$ for main-sequence stars (over at least 3 orders of magnitude) indicates that a lower $L_x$ limit in close-orbit planet systems may have little effect on the mean emission of a population. 

We tentatively examine the hypothesis that the observed $L_x$-mass correlation is a direct consequence of magnetic field interaction between the planets and their host stars that results in the dissipation of energy, some of which results in enhanced X-ray emission from the coronal plasma. No simple predictive model exists for the contribution to coronal emission in this case. \citet{cuntz2000} make a first-principles estimate of the power of magnetic field interaction between a planet and star, including reconnection due to both the relative motion of a planet and the coronal field and the `steady-state' (i.e. in the case of full stellar synchronization with the planet orbit), due to photospheric motions, field tangling and subsequent reconnection. To zeroth-order they suggest that dissipated power scales as $\sim B_p^{1/3}$, where $B_p$ is magnetic field strength at the planet's poles -  with dependencies on stellar field, orbital distance, planet size and factors associated with the magnetospheric radius (interaction point).  However, it has been pointed out that magnetic reconnection energy release would  not be expected to be sufficient to explain present observational constraints on `hot-spots' in stellar chromospheres due to planet-star interaction \citep{lanza2009b}. \citet{lanza2009b} further suggests that planets can cause a release of built up coronal field energy by dissipating field helicity. In this case the dissipated power scales as $\sim B_p^{2/3}$ (plus of course a dependency on the stellar magnetic field strength, orbital distance and other physical parameters). This does however assume that magnetic pressure dominates over thermal and ram pressure in the stellar coronal pressure, which may not be valid \citep{petrinec1997}. 

Recent efforts at magneto-hydrodynamical (MHD) simulation of star-planet interaction also suggest that total X-ray luminosity can increase \citep{cohen2009}. However, in this case it is a consequence of localized increases in coronal plasma density due to the interaction, since thermal bremsstrahlung emission scales as plasma density squared. In such models the net X-ray emission of a star can be enhanced by 10-30\% and possibly by even an order of magnitude due to the presence of a close-in planet. This is certainly consistent with our results, however the simulations span too limited a range of situations at present to allow us to make a quantitative comparison.

Tidal locking \citep{goldreich1966} of inner planets may reduce convectively-driven dynamos (e.g. \citet{griessmeier2004}), however induced currents in planetary conductive interiors would sustain planet-star interaction, albeit with a power dissipation efficiency lower by a few orders of magnitude \citep{lanza2009b}.

If we naively assume that the observed correlation of $L_x$ with planet mass in our inner subsample is directly proportional to dissipated power through magnetic field interaction we can estimate (for example) the ratio of planetary magnetic field strength between 10 M$_J$  and 1 M$_J$  planets. If dissipated power scales as $B_p^{1/3}$ then $(L_x^{10}/L_x^1)^3 \approx B_p^{10}/B_p^1\approx 59^{+19}_{-14}$, implicitly ignoring all errors and scatter due to dependencies on stellar field strengths, age, orbital distance and other parameters. If the power scaled as $B_p^{2/3}$ we then estimate $(L_x^{10}/L_x^1)^{3/2} \approx B_p^{10}/B_p^1\approx 8^{+1}_{-1}$. If the dissipated power is not directly proportional to the observed $L_x$ variation with planet mass but is rather assumed to go into raising the coronal temperature then since $L_x \propto T^{1/2}$ to first-order for a thermal plasma we then estimate (assuming dissipation proportional to $B_p^{2/3}$) that $B_p^{10}/B_p^1\sim 64$. 

Recent geo-dynamical modelling work on the origins of magnetic fields has included a proposed scaling law that successfully predicts magnetic field strengths from Earth mass planets to rapidly rotating low-mass stars \citep{christensen2009}. This scaling law predicts $B_p^{10}/B_p^{1} = 12^{+9}_{-3}$ ($1\sigma$ uncertainties), which would be in close agreement with our estimate assuming a linear $L_x$ relation to dissipated power going as $B_p^{2/3}$. Interestingly, our higher estimates for this ratio ($\sim 60$) can probably be ruled out, since measured old M-dwarf and T-Tauri star field strengths \citep{christensen2009} would then be comparable to that of $10 $ M$_J$ planets.

\section{Discussion}

Using a sample of exoplanet host stars we have examined whether their X-ray emission shows any evidence for energy dissipation due to planet-star interaction. While we find no statistical evidence of enhanced emission for systems harboring the closest-orbit planets ($<0.15$ AU periastron distances), we do find evidence for a positive correlation between X-ray luminosity and $M_p\sin i$ that is not seen in systems with more distant known planetary companions. Our analyses include a full treatment of censored data (with detection upper limits) using the tools of survival analysis. We suggest that the observed correlation may represent a lower-limit or floor to emission, generated by dissipative (or coronal density enhancement) processes of planet-star magnetic field interaction. Intriguingly, assuming the favored model for dissipated power in planet-star magnetic interaction - scaling as $B_p^{2/3}$ - we can crudely estimate the ratio of magnetic field strengths of 10 and 1 M$_J$ planets as $\sim 8^{+1}_{-1}$, which is in remarkably close agreement with the predictions of geo-dynamical modeling that successfully matches data ranging from planets to stars \citep{christensen2009} and suggests a ratio $\sim 12^{+9}_{-3}$. In making this estimate we relegate several other physically important parameters (orbital radius, stellar magnetic field, coronal magnetic field rotation) to the role of producing scatter - and assume that we nonetheless measure a mean relationship between X-ray  luminosity and projected planet mass. Further investigation of this is clearly needed.

We have examined possible observational systematics and selection biases that could produce the observed correlation, or confuse its interpretation. While there is evidence for observational bias due to RV detection sensitivity that results in the lowest projected mass planets in our sample (all with $<0.04$ M$_J$) having stellar parents with lower $L_x$ (all $<5 \times 10^{27}$ erg s$^{-1}$), there is no corresponding evidence for systems with higher mass planets ($>0.1$ M$_J$).  We have also examined the relationship of published data on stellar rotation ($v_*\sin i$) to both X-ray emission and planet mass. While we see evidence for the stellar hosts of lower mass, close-in, planets ($M_p\sin i <0.04$ M$_J$) to have systematically slower rotation ($\sim 1$ km s$^{-1}$), there is no clear evidence for any systematic trend in other systems. Again, this indicates - not surprisingly - that the slowest rotating stars allow for the most sensitive RV measurements and therefore enable the detection of the lowest mass planets. We have also examined the empirically derived relationship between stellar X-ray luminosity and rotation ($L_x \propto (v_*\sin i)^2$ erg s$^{-1}$) and observed luminosities and find no obvious indicators that the distribution would influence an apparent $L_x-M_p\sin i$ correlation. We also conclude that planet atmospheric erosion by high-energy photons does not appear to modify populations in the sense that would explain our observed correlation.

Inclination effects could conceivably bias the measurement of planet-stimulated coronal X-ray flux towards
systems of low inclination (\S4.1), with correspondingly larger projected masses. Transiting systems could help resolve this question, however at this time only one system with an observed planetary transit is in the X-ray detected sample.

\section{Conclusions}

The data presented here indicate that X-ray emission from planet-hosting stars may offer a probe of planet-star interactions, and conceivably a probe of the inner workings of close-orbit giant worlds through their convective-dynamo generated magnetic fields. The present sample is however small and more sensitive, targeted, X-ray data on planet hosting stars is needed. This data must also fully sample emission during the entire planetary orbital period in order to avoid biasing any measurement. Current observational programs are beginning to fill this gap (e.g. \citet{poppenhaeger2010, sanz-forcada2010}), but sample selection must also be made in a uniform and unbiased fashion in order for these data to be of greatest utility. Given the inherent biases in RV and transit detection of planets this will be particularly critical.

As present works show \citep{shkolnik2008,walker2008}, it is clear that planet-star interaction in specific cases can directly affect observables at a variety of wavelengths, and that the detailed study of individual objects will also be extremely valuable. In particular, obtaining sensitive time-series X-ray data of close-orbit, {\em transiting} planet systems over multiple orbital periods and epochs may help disentangle inclination effects, magnetic field dissipation or plasma density enhancement mechanisms, and intrinsic stellar variations.

\acknowledgments

This work was supported by a NASA Astrophysics Data Analysis grant NNX08AJ54G. This research has made use of data obtained from the High Energy Astrophysics Science Archive Research Center (HEASARC), provided by NASA's Goddard Space Flight Center. C.A.S. thanks the anonymous referees for their helpful comments at various stages during this work.

\clearpage


\clearpage

\begin{figure}
\epsscale{0.5}
\plotone{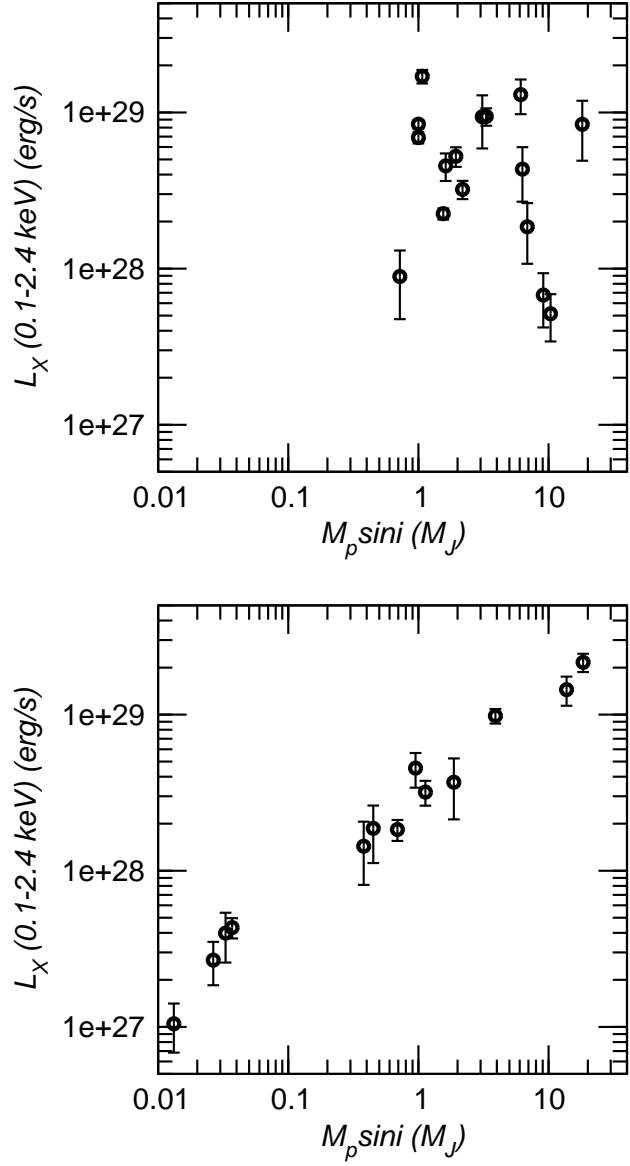}
\caption{X-ray luminosity is plotted versus planet mass ($M_p\sin i$). Upper panel is outer planet ($d_{peri}>0.15$ AU) subsample, lower panel is inner planet ($d_{peri}<0.15$ AU) subsample. $1\sigma$ errors on $L_x$ are derived from Poisson uncertainties on photon counts, including background subtraction. \label{fig3}}
\end{figure}

\clearpage

\begin{figure}
\epsscale{0.5}
\plotone{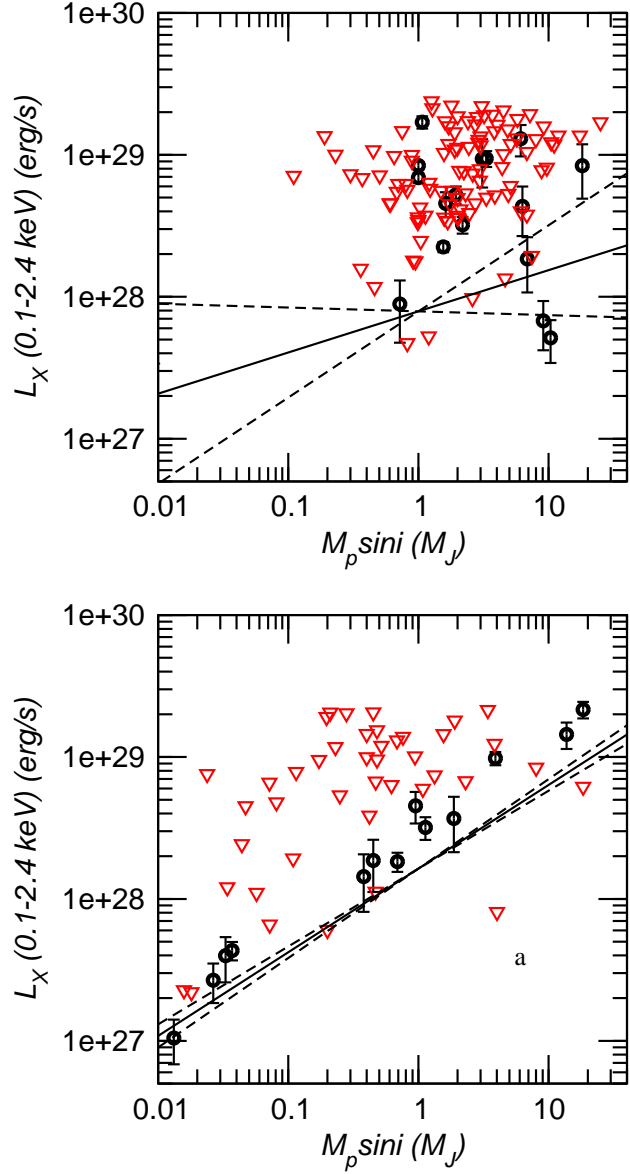}
\caption{Measured X-ray luminosities (heavy circles) and estimated upper limits (open triangles) to X-ray luminosities are plotted versus $M_p \sin i$ (M$_J$) for systems with exoplanet periastron distances $> 0.15$ AU (upper panel) and $< 0.15$ AU (lower panel).  The system GL86 is labeled as `a' in the lower panel and discussed in the text. Best fit power-laws from BJ regression analyses (allowing for censored data) are plotted as solid lines, dashed lines indicate the range of formal uncertainty in slope. The fit to the outer planet subsample (upper panel) is consistent with rank correlation analyses (\S 3) that indicate no significant correlation in that dataset.}
\end{figure}

\clearpage

\begin{figure}[h]
\epsscale{1.0}
\plotone{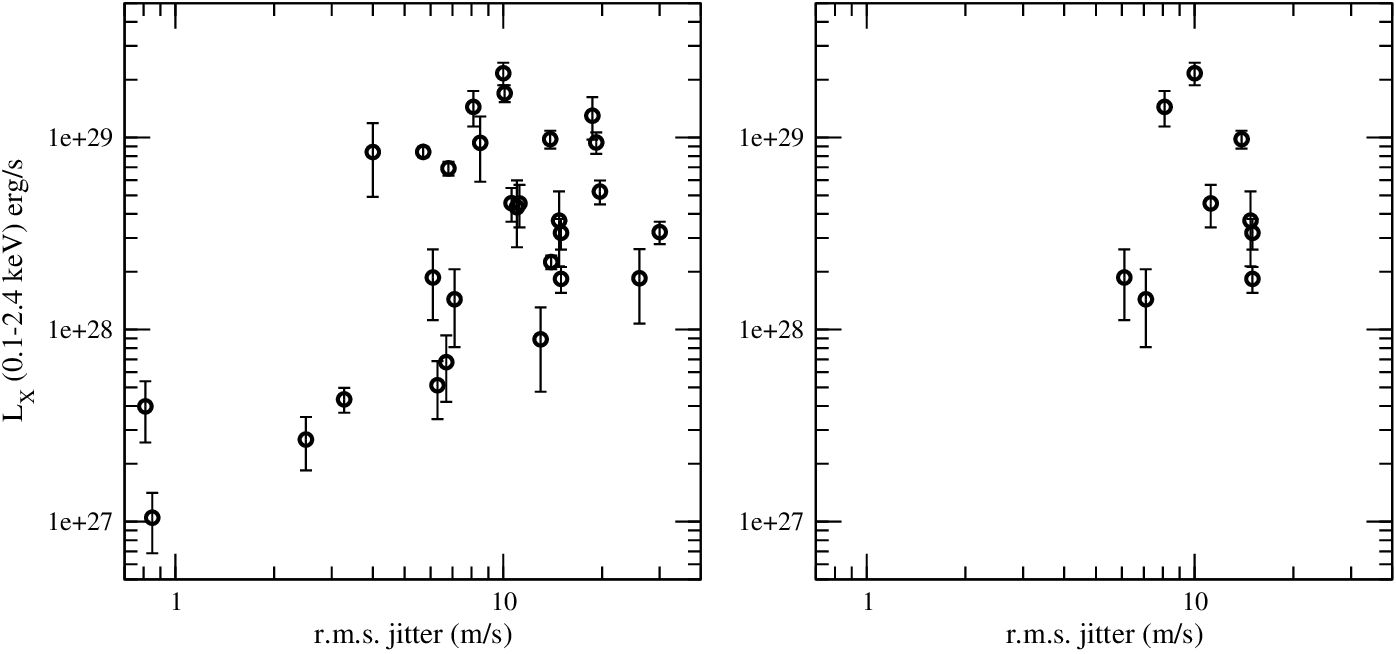}
\caption{$L_x$ is plotted versus the residual stellar jitter for the complete planet sample with X-ray detections (left panel) and inner planet subsample with $M_p\sin i > 0.1$ M$_{J}$ (right panel).
 \label{fig4}}
\end{figure}

\clearpage

\begin{figure}[h]
\epsscale{0.7}
\plotone{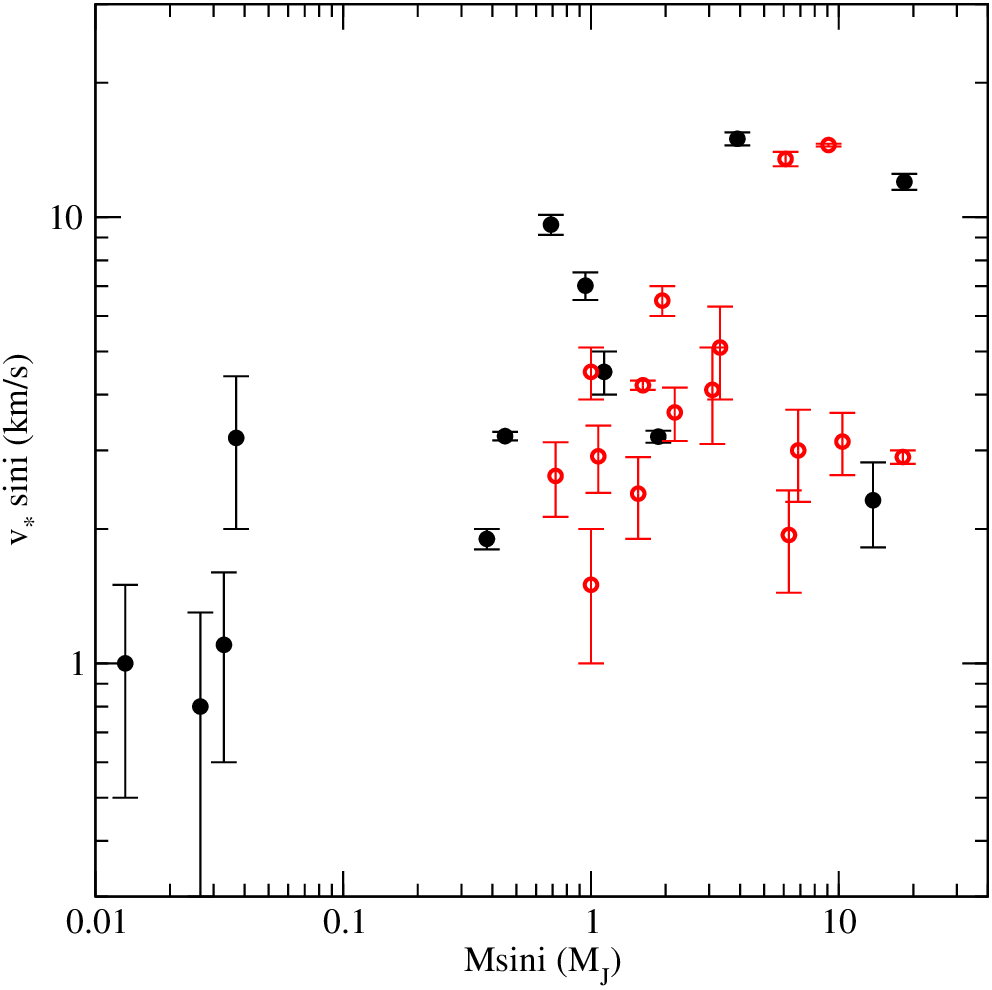}
\caption{Estimated stellar projected rotation velocities ($v_*\sin i$) are plotted versus $M_p\sin i$ for both inner (filled symbols) and outer (open symbols) planet subsamples. 
 \label{fig4}}
\end{figure}

\clearpage

\begin{figure}[h]
\epsscale{0.7}
\plotone{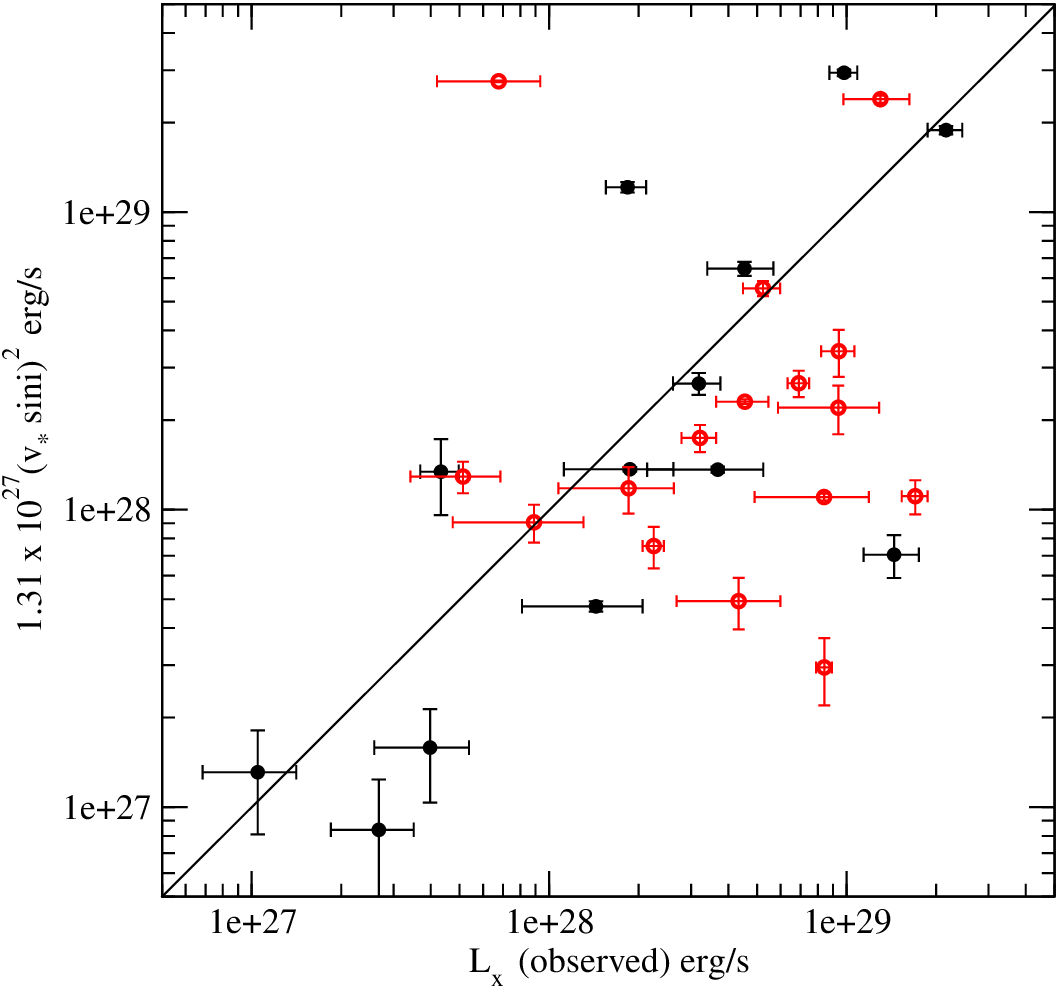}
\caption{X-ray luminosity predicted from stellar rotation is plotted versus the observed luminosity for both the inner (filled symbols) and outer (open symbols)  planet subsamples used here. Solid line indicates 1:1 matching of the luminosities.
 \label{fig4}}
\end{figure}

\clearpage

\begin{figure}[h]
\epsscale{0.9}
\plotone{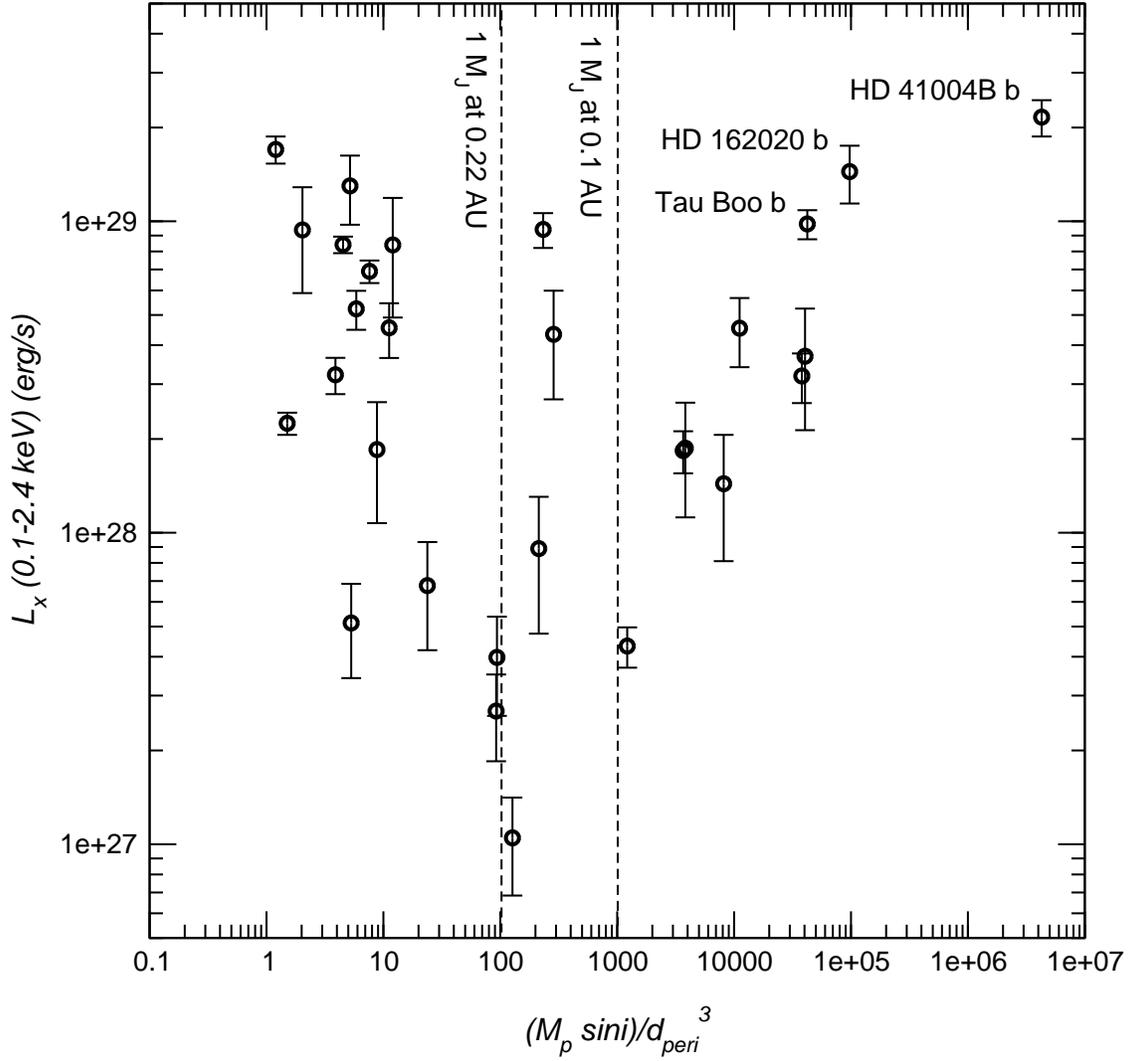}
\caption{The measured X-ray luminosity for each system is plotted versus a factor proportional to the height of the tidal bulge raised in the stellar atmosphere by the innermost exoplanet. Vertical dashed lines illustrate planet masses and orbital radii combinations that correspond to particular tidal factors. We consider factors $>1000$ to indicate a physically significant tidal interaction. Three systems with the largest tides are labeled and correspond to planets of mass $M_p\sin i=3.9$ M$_J$ (Tau Boo), $M_p\sin i=13.75$ M$_J$ (HD 16202 b), and $M_p\sin i=18.4$ M$_J$ (HD 41004B b).
 \label{fig9}}
\end{figure}









\clearpage

\begin{deluxetable}{lllllll} 
\tabletypesize{\small}
\tablecolumns{7} 
\tablewidth{0pc} 
\tablecaption{Planet and stellar properties of exoplanet sample} 
\tablehead{ Planet ID & $M_p\sin i$& Semi-major & Periastron & $L_x$ (0.1-2.4 keV)  & Poisson ($1\sigma$) & r.m.s.  \\ 
\colhead{}  &  (M$_J$) & axis (AU) & distance (AU) & erg s$^{-1}$ & error on $L_x$ & jitter (ms$^{-1}$) \\
}
\startdata 
HD 40307 b &	0.0132 &	0.047	& 0.047	&$1.05\times 10^{27}$	&$3.64\times 10^{26}$	&0.85\\
HD 285968 b& 	0.0265 &	0.066 &	0.066 &	$2.68\times 10^{27}$ &	$8.30\times 10^{26}$ &	2.5\\
HD 69830 b &	0.033 &	0.0785 &	0.0707 &	$3.98\times 10^{27}$ &	$1.40\times 10^{27}$ &	0.81\\
GJ 674 b & 0.037 &	0.039 &	0.0312 &	$4.33\times 10^{27}$ &	$6.41\times 10^{26}$ &	3.27\\
HD 63454 b &	0.38 &	0.036 &	0.036 &	$1.44\times 10^{28}$ &	$6.26\times 10^{27}$ &	7.1\\
HD 102195 b &	0.45 &	0.049 &	0.049 &	$1.87\times 10^{28}$ &	$7.47\times 10^{27}$ &	6.1\\
Ups And b &	0.69 &	0.059 &	0.0573 &	$1.83\times 10^{28}$ &	$2.82 \times 10^{27}$ &	15\\
HD 192263 b &	0.72 &	0.15 &	0.15 &	$8.89\times 10^{27}$ &	$4.15\times 10^{27}$ &	13\\
HD 179949 b &	0.95 &	0.045 &	0.04401 &	 $4.54\times 10^{28}$	 &  $1.13\times 10^{28}$ & 11.2\\
HD 147513 b &	1 &	1.26 &	0.6048 &	$8.41\times 10^{28}$ & $5.18\times 10^{27}$ &	5.7\\
HD 150706 b &	1 &	0.82 &	0.5084 &	$6.90\times 10^{28}$ & $5.75\times 10^{27}$ &	6.8\\
HD 20367 b &	1.07 &	1.25 &	0.9625 &	$1.70\times 10^{29}$ &	$1.70\times 10^{28}$ &	10.1\\
HD 189733 b &	1.13 &	0.03099 &	 0.03099 & $3.19\times 10^{28}$ &	$5.79\times 10^{27}$ &	15\\
E. Eridani b &	1.55 &	3.39 &	1.01022 &	 $2.25\times 10^{28}$ &	$1.83\times 10^{27}$&   14\\
HD 142415 b &	1.62 &	1.05 &	0.525 &	$4.55\times 10^{28}$ &	$9.10\times 10^{27}$ &	10.6\\
HD 73256 b &	1.87 &	0.037 &	0.03589 &	 $3.69\times 10^{28}$ &	$1.55\times 10^{28}$ &	14.8\\
HR 810 b &	1.94 &	0.91 &	0.6916 &	$5.23\times 10^{28}$ &	$7.47\times 10^{27}$ &	19.7\\
HD 128311 b &	2.18 &	1.099 &	0.8243 &	$3.21\times 10^{28}$ &	$4.29\times 10^{27}$ &	30\\
HD 221287 b &	3.09 &	1.25 &	1.15 &	$9.37\times 10^{28}$ &	$3.49\times 10^{28}$ &	8.5\\
GJ 3021 b	 & 3.32 &	0.49	& 0.24255	 & $9.42\times 10^{28}$ &	$1.21\times 10^{28}$ &	19.2\\
Tau Boo b &	3.9 & 	0.046 &	0.0452 &	$9.80\times 10^{28}$ &	$1.05\times 10^{28}$ &	13.9\\
HD 70573 b &	6.1 &	 1.76 &	1.056 &	$1.30\times 10^{29}$ &	$3.25\times 10^{28}$ &	18.7\\
HD 178911B b &	6.292 &	0.32 &	0.2802 &	$4.33\times 10^{28}$ & $1.65\times 10^{28}$ &	11\\
HD 81040 b &	6.86 &	1.94 &	0.91956 & $1.85\times 10^{28}$ &	$7.77\times 10^{27}$ &	26\\
HD 33564 b &	9.1 &	 1.1 &	0.726 &	$6.76\times 10^{27}$ &	$2.57\times 10^{27}$ &	6.7\\
HD 39091 b &	10.35 &	3.29 &	1.2502 &	$5.13\times 10^{27}$ & $1.72\times 10^{27}$ &	6.3\\
HD 162020 b &	 13.75 &	0.072 &	0.05205 &	 $1.44\times 10^{29}$ &	$3.04\times 10^{28}$ & 8.1\\
HD 131664 b &	18.15 &	3.17 &	1.14754 &	 $8.40\times 10^{28}$ & $3.49\times 10^{28}$ &	4\\
HD 41004B b &	18.4 &	0.0177 &	0.0163 &	$2.16\times 10^{29}$ &	$2.88\times 10^{28}$ &	10\\
\enddata 
\tablecomments{The primary data used in our analyses is presented. Columns are: 1) name of planet candidate, 2) projected mass from radial velocity measurements in units of Jupiter mass, taken as the most recently cited/trusted value \citep{schneider2010}, 3) semi-major axis of planet orbit, taken as the most recently cited/trusted value, 4) periastron distance (closest approach) of planet evaluated as $a(1-e)$, 5) X-ray luminosity derived from ROSAT catalogues in the 0.1-2.4 keV band, 6) $1-\sigma$ error on X-ray luminosity derived from the quoted count rate error in the ROSAT catalogs that includes the Poisson uncertainty in total count rate and estimated background count rate. Errors due to the finite precision of the Hipparcos astrometric catalog are not included. 7) The r.m.s. velocity error or ÒjitterÓ from the residuals of the best-fit planetary orbit to radial velocity data, obtained from the most recent, primary publication listed in the online exoplanet catalog \citep{schneider2010}.
 }
\end{deluxetable} 


\begin{deluxetable}{llll} 
\tabletypesize{\small}
\tablecolumns{3} 
\tablewidth{0pc} 
\tablecaption{K-M estimator tests of luminosity detection and upper limit distributions  for stellar samples (units of $\log_{10}L_x$ erg s $^{-1}$, 0.1-2.4 keV band). \label{tbl-2}}
\tablehead{Sample & 75th percentile & 50th percentile & 25th percentile \\}
\startdata 
Inner planet systems ($d_{peri}<0.15$ AU) & 21.169 & 27.549 & 28.265 \\
Outer planet systems ($d_{peri}>0.15$ AU) & 27.683 & 27.805 & 28.313 \\
\tableline
Inner sample mean & 27.735 $\pm 0.144$ &  - &- \\
Outer sample mean & 28.056 $\pm 0.078$ & - & -\\
\enddata
\end{deluxetable}

\begin{deluxetable}{llll} 
\tabletypesize{\small}
\tablecolumns{3} 
\tablewidth{0pc} 
\tablecaption{Buckley-James regression and EM algorithm regression results for $\log_{10} L_x$ versus $\log_{10} M_p\sin i$  (luminosity in 0.1-2.4 keV band). \label{tbl-2}}
\tablehead{Parameter & Inner planet & Inner planet & Outer planet   \\
 & subsample & subsample (GL 86 removed) & subsample \\}
\startdata 
BJ regression: &  &  & \\
 Intercept & 28.21  & 28.36 & $27.90$ \\
Slope & $0.59 \pm 0.04$  & $0.63 \pm 0.03$ & $0.29 \pm 0.32$\\
Standard deviation & 0.15 & 0.12 & 0.52 \\  
\tableline
EM algorithm: & & & \\
Intercept & $28.14 \pm 0.10$ & $28.36\pm 0.05$  & $27.91\pm 0.16$ \\
Slope     &  $0.61 \pm 0.09$    &  $0.63\pm 0.04$  & $0.30 \pm 0.21$ \\
Standard deviation & 0.40  &  0.19   & 0.53 \\      
\enddata
\end{deluxetable}

\end{document}